\documentclass[aps,twocolumn,groupeaddress,prl]{revtex4}
\usepackage{amssymb}
\usepackage{amsmath}
\usepackage{graphicx}

\def\kb{k_{B}}
\def\te{T_{e}}

\def\sit{S_{t}}
\def\sib{S_{b}}
\def\six{S_{tb}}
\def\vtl{V_{tl}}
\def\vtc{V_{tc}}
\def\vtr{V_{tr}}
\def\vbl{V_{bl}}
\def\vbc{V_{bc}}
\def\vbr{V_{br}}
\def\vl{V_{l}}
\def\vr{V_{r}}
\def\It{I_{t}}
\def\Ib{I_{b}}
\def\vt{V_{t}}
\def\vb{V_{b}}
\def\gt{g_{t}}
\def\gb{g_{b}}
\def\ohm{\mathrm{\Omega}}
\def\kohm{\mathrm{k\Omega}}

\def\muV{\mu\mathrm{V}}
\def\mueV{\mu\mathrm{eV}}
\def\MHz{\mathrm{MHz}}

\def\mk{\mathrm{mK}}
\def\Hethree{^3\mathrm{He}}
\def\Mmat{\mathcal{M}}
\def\Jtr{J^{{tr}}}
\def\Jbr{J^{{br}}}
\def\brho{\boldsymbol{\rho}}
\def\brhobar{\boldsymbol{\bar{\rho}}}
\def\rate{W}

\begin{document}

\title{Tunable Noise Cross-Correlations in a Double Quantum Dot}
\author{D.\ T.\ McClure, L.\ DiCarlo, Y.\ Zhang, H.-A.\ Engel and C.\ M.\ Marcus}
\affiliation{Department of Physics, Harvard University, Cambridge,
Massachusetts 02138, USA}
\author{M.\ P.\ Hanson and A.\ C.\ Gossard}
\affiliation{Department of Materials, University of California,
Santa Barbara, California, 93106, USA}
\date{\today}

\begin{abstract}
We report measurements of the cross-correlation between current noise fluctuations
in two capacitively coupled quantum dots in the Coulomb blockade regime. The sign
of the cross-spectral density is found to be tunable by gate voltage and
source-drain bias. Good agreement is found with a model of sequential tunneling
through the dots in the presence of inter-dot capacitive coupling.
\end{abstract}

\maketitle

Current noise cross-correlation in mesoscopic electronics is of broad interest
because it is sensitive to quantum indistinguishability and
interactions~\cite{blanter00-05,martin03,samuelsson04,beenakker04,lebedev05}. In
the absence of interactions, Pauli exclusion makes the cross-correlation between
any two outgoing currents negative, regardless of device geometry, temperature,
and bias voltage~\cite{buttiker90-92}. Experiments to date, notably the electronic
versions of the Hanbury Brown-Twiss
experiment~\cite{henny99,oliver99,oberholzer00}, have shown good agreement with
this theoretical prediction.

However, the negative sign of the cross-correlation rests on a number of
assumptions: a low-impedance external circuit, low measurement frequency, and
vanishing interactions. Recent
work~\cite{texier00,buttiker03,wu05,oberholzer06,rychkov06} explores the
possibility of controlling the sign of cross-correlations by relaxing one or more
of these conditions. For example, if the first assumption is relaxed, sign
reversal can result from strong inelastic scattering, as theoretically
predicted~\cite{texier00} and experimentally demonstrated~\cite{oberholzer06}
using a voltage probe. Furthermore, sign reversal can also occur in hybrid
superconductor-normal systems when non-local Andreev reflection dominates over
normal scattering, as was shown theoretically in
Refs.~\cite{anantram96,martin96,torres99}.

\begin{figure}[t!]
\center \label{figure1}
\includegraphics[width=3.25in]{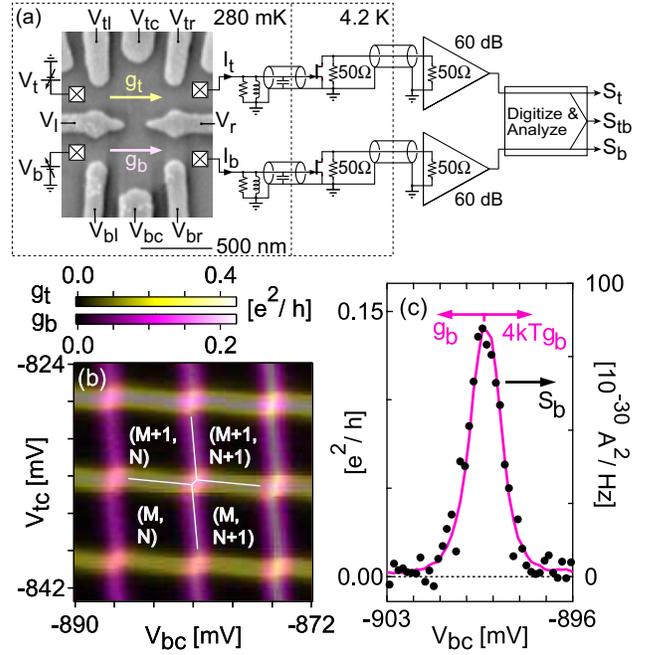}
\caption{\footnotesize{(a) Scanning electron micrograph of the double-dot device,
and equivalent circuit at $2~\MHz$ of the noise detection system measuring the
power spectral densities and cross spectral density of fluctuations in currents
$\It$ and $\Ib$. (b) Differential conductances $\gt$ (yellow) and $\gb$ (magenta)
as a function of $\vtc$ and $\vbc$ over a few Coulomb blockade peaks in each dot,
at $\vt=\vb=0$. Black regions correspond to well-defined charge states in the
double-dot system. Superimposed white lines indicate the honeycomb structure
resulting from the finite inter-dot capacitive coupling. (c) Zero-bias (thermal)
noise $\sib$ (black dots, right axis), conductance $\gb$ (magenta curve, left
axis), and calculated $4\kb\te\gb$ (magenta curve, right axis) as a function of
gate voltage $\vbc$.}}
\end{figure}

In this Letter, we report gate-controlled sign reversal of noise cross-correlation
due to strong Coulomb interaction in a capacitively coupled double quantum dot.
Previously, super-Poissonian current noise was observed in mesoscopic tunnel
junctions~\cite{chen05} and silicon MOSFETs at low transmission~\cite{safonov03}
and attributed to Coulomb interaction between localized states mediating
transport. In contrast to devices with uncontrolled localized states, the present
structure allows full control of Coulomb-induced correlation via the electrostatic
gates that define the two dots. A model of sequential tunneling through a single
level within each dot that includes capacitive coupling yields good agreement
with
the experimental results. This study provides an intuitive picture of how Coulomb
interaction influences noise cross-correlation, taking advantage of a
well-controlled device geometry.

The four-terminal double-dot device is defined by top gates on the surface of a
$\mathrm{GaAs}/\mathrm{Al}_{0.3}\mathrm{Ga}_{0.7}\mathrm{As}$ heterostructure
grown by molecular beam epitaxy [see micrograph in Fig.~1(a)]. The two-dimensional
electron gas $100~\mathrm{nm}$ below the surface has a density of
$2\times10^{11}~\mathrm{cm}^{-2}$ and mobility $2\times10^5~
\mathrm{cm}^2/\mathrm{Vs}$. Throughout this experiment, gate voltages
$\vl=\vr=-1420~\mathrm{mV}$ fully deplete the central point contact such that the
dots are coupled only capacitively, not via tunneling. Gate voltages $\vtl$
($\vbl$) and  $\vtr$ ($\vbr$) control the tunnel barrier between the top (bottom)
dot and its left and right leads. Plunger gate voltage $\vtc$ ($\vbc$) controls
the electron number $M$ ($N$) in the top (bottom) dot; for this experiment $M\sim
N\sim 100$. The lithographic area of each dot
is $0.15~\mu\mathrm{m}^2$. We estimate a mean level spacing of each dot $\Delta_{t(b)}\approx
70~\mueV$, assuming $\sim 100~\mathrm{nm}$ of lateral depletion
around the gates.

Measurements are carried out in a $\Hethree$ cryostat equipped with a two-channel
noise measurement system~\cite{techniques}, shown schematically in Fig.~1(a).
A voltage bias $\vt$ ($\vb$) is applied to the left lead of the top
(bottom) dot, with right leads held at ground. Separate resistor-inductor-capacitor resonators
($R=5~\kohm,~L=66~\mu\mathrm{H},~C=96~\mathrm{pF}$) convert fluctuations in
currents $I_t$ and $I_b$ through the top and bottom dots around $2~\MHz$ into
voltage fluctuations on gates of high electron mobility transistors (HEMTs) at
4.2~K, which in turn convert these into current fluctuations in two $50~\ohm$
coaxial lines extending to room temperature, where further amplification is
performed. These signals are then simultaneously digitized at $10~\MHz$, their
fast Fourier transforms calculated, and the current noise power spectral densities
$\sit, \sib$ and cross-spectral density $\six$ extracted. The overall gain of each
amplification channel and the base electron temperature $\te=280~\mk$ are
calibrated \textit{in situ} using Johnson-noise thermometry at base temperature
and 1.6~K with the device configured as two point contacts~\cite{techniques}.
Differential conductance $\gt$ ($\gb$) through the top (bottom) dot is measured
using standard lock-in technique with an applied excitation of
25~(30)~$\muV_{\mathrm{rms}}$ at 677~(1000)~Hz. Ohmic contact resistances of
roughly a few $\kohm$, much smaller than the dot resistances, are not subtracted.

Superposed top and bottom dot conductances $\gt$ and $\gb$ as a function of
plunger voltages $\vtc$ and $\vbc$ form the characteristic double-dot honeycomb
pattern~\cite{vanderwiel03,chan02}, with dark regions corresponding to
well-defined electron number in each dot, denoted $(M,N)$ (first index for top
dot), as shown in Fig.~1(b). Horizontal (vertical) features in $\gt$ ($\gb$) are
Coulomb blockade (CB) conductance peaks~\cite{kouwenhoven97}, across which $M$
$(N)$ increases by one as $\vtc$ $(\vbc)$ is raised. The distance between triple
points, i.e., the length of the short edge of the hexagon, provides a measure of
the mutual charging energy $U$ due to inter-dot capacitive coupling. By comparing
this distance to the CB peak spacing, and using the single-dot charging energy
$E_C=600~\mu\mathrm{eV}$ extracted from finite bias CB diamonds (not shown), we
estimate $U\approx60~\mu\mathrm{eV}$~\cite{chan02}. We refer to the midpoint of
the short edge of the hexagon, midway between triple points, as the ``honeycomb
vertex.'' Current noise $\sib$ and conductance $\gb$, measured simultaneously
around zero-bias, over a CB peak in the bottom dot (with the top dot in a CB
valley) are shown in Fig.~1(c). Good agreement between the measured $\sib$ and the
Johnson-Nyquist thermal noise value $4\kb\te\gb$ is observed.

\begin{figure}[t]
\center \label{figure2}
\includegraphics[width=3.25in]{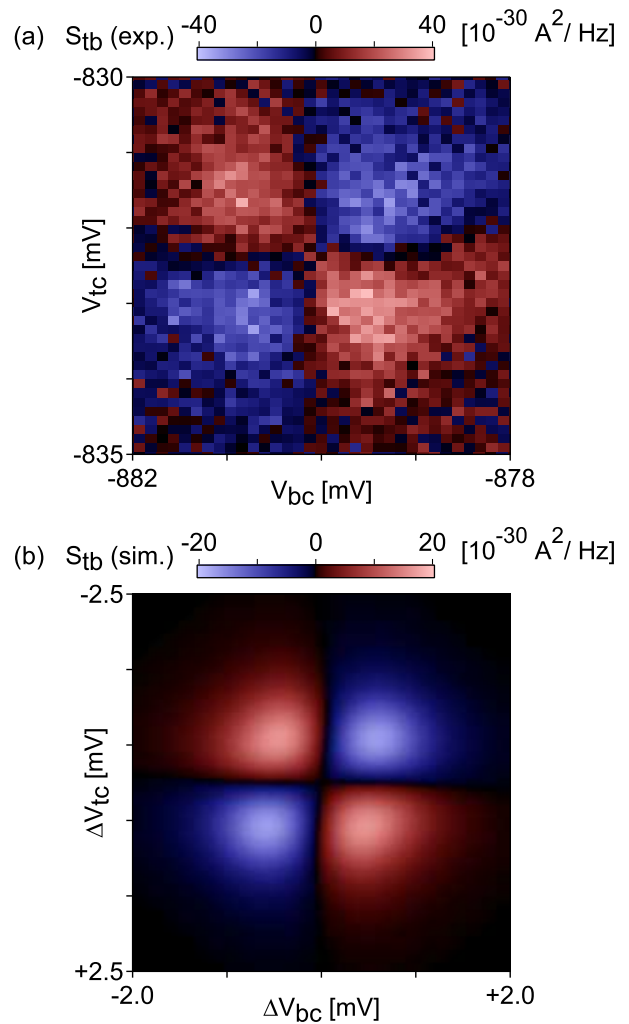}
\caption{\footnotesize{Measured (a) and simulated (b) cross-spectral density
$\six$ near a honeycomb vertex, with applied bias $\vt=\vb=-100~\muV$
$(e|V_{t(b)}|\approx 4\kb\te\approx E_{C}/6)$. Blue regions (lower-left and
upper-right) indicate negative cross-correlation, while red regions indicate
positive cross-correlation.}}
\end{figure}

Turning now to finite-bias noise measurements, Fig.~2(a) shows the measured noise
cross-correlation $\six$ as a function of plunger gate voltages $\vtc$ and $\vbc$
in the vicinity of a honeycomb vertex, with voltage bias of $-100~\muV$ applied to
both dots. The plot reveals a characteristic quadrupole pattern of noise
cross-correlation centered on the honeycomb vertex, comprising both negative and
positive cross-correlation regions. The symmetry of this pattern is found to
depend rather sensitively on balancing each dot's left and right tunnel barriers.

\begin{figure}[t]
\center \label{figure3}
\includegraphics[width=3.25in]{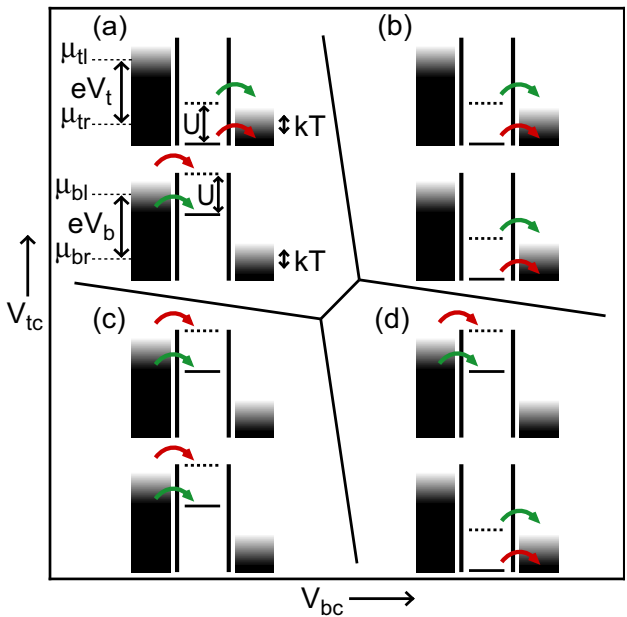}
\caption{\footnotesize{Energy level diagrams in the vicinity of a honeycomb
vertex, with biases $V_{t(b)}=-100~\muV$. (The various energies are shown roughly
to scale.) 
The solid horizontal line in the top (bottom) dot represents the energy $E_{t(b)}$ required
to add electron $M+1$ $(N+1)$ when the bottom (top) dot has $N$ $(M)$ electrons.
The dashed horizontal line, higher than the solid line by $U$, represents
$E_{t(b)}$ when the bottom (top) dot has $N+1$ $(M+1)$ electrons. In each dot, the
rate of either tunneling-in from the left or tunneling-out to the right is
significantly affected by this difference in the energy level, taking on either a
slow value (red arrow) or a fast value (green arrow) depending on the electron
number in the other dot. In (a) and (d), where the occurrence of each
$U$-sensitive process enhances the rate of the other, we find positive
cross-correlation. In (b) and (c), where the occurrence of each $U$-sensitive
process suppresses the rate of the other, we find negative cross-correlation.}}
\end{figure}

To better understand this experimental result, we model the system as single-level
dots capacitively coupled by a mutual charging energy $U$, each with weak
tunneling to the leads. The energy required to add electron $M+1$ to the top dot
depends on the two plunger gate voltages as well as the electron number
$n\in\{N,N+1\}$ on the bottom dot: $E_t=\alpha_t\vtc+\beta_t\vbc+U\cdot
n+\mathrm{const.}$, where lever arms $\alpha_t$ and $\beta_t$ are obtained from
the honeycomb diagram in Fig.~1(b)~\cite{vanderwiel03} and the measured $E_C$. The
energy $E_b$ to add electron $N+1$ to the bottom dot is given by an analogous
formula. Occupation probabilities for charge states $(M,N)$, $(M+1,N)$, $(M,N+1)$,
and $(M+1,N+1)$ are given by the diagonal elements of the density matrix,
$\brho=\left(\rho_{00},\rho_{10},\rho_{01},\rho_{11} \right)^{T}$. The time
evolution of $\brho$ is given by a master equation $d\brho/dt= \mathcal{M}\brho$,
where
\begin{equation}
\mathcal{M}=\left(
\begin{array}{cccc}
-\rate_{00}^{\mathrm{out}} & \rate_{00\leftarrow10} &\rate_{00\leftarrow01} & 0 \\
\rate_{10\leftarrow00} & -\rate_{10}^{\mathrm{out}} & 0 & \rate_{01\leftarrow11}\\
\rate_{01\leftarrow00} & 0 & -\rate_{01}^{\mathrm{out}} & \rate_{10\leftarrow11}\\
0 & \rate_{11\leftarrow10} & \rate_{11\leftarrow01} & -\rate_{11}^{\mathrm{out}}\\
\end{array}
\right).
\end{equation}
Each diagonal element of $\Mmat$ gives the total loss rate for the corresponding
state, with $\rate_\alpha^\mathrm{out}=\sum_\beta \rate_{\beta\leftarrow\alpha}$.
Off-diagonal terms give the total rate for transitions between two states. For
example, $\rate_{10\leftarrow
00}=\rate^{l}_{10\leftarrow00}+\rate^{r}_{10\leftarrow00}$ is the total tunneling
rate into $(M+1,N)$ from $(M,N)$, containing tunneling contributions from the
top-left and top-right leads.

Rates for tunneling-in and tunneling-out between a dot and either of its leads
$i\in\{tl,tr,bl,br\}$ depend on both the transparency $\Gamma^i$ of the tunnel
barrier to lead $i$ and the Fermi function
$f_i(\epsilon)=\left[1+\mathrm{exp}\left\{(\epsilon-\mu_i)/\kb\te\right\}\right]^{-1}$
evaluated at the value at $\epsilon=E_{t(b)}$, where $\mu_i$ is the chemical
potential in lead $i$. For example, the rates for tunneling into and out of the
top dot from/to the left lead are given by $\rate^{l}_{10\leftarrow 00}=
\Gamma^{lt} f_{lt}(E_t)$ and $\rate^{l}_{00\leftarrow
10}=\Gamma^{lt}\left[1-f_{lt}(E_t)\right]$, respectively. As $E_{t}$ is lowered
across $\mu_{lt}$, $\rate^{l}_{10\leftarrow 00}$ increases from 0 to $\Gamma^{lt}$
over a range of a few $\kb\te$, while $\rate^{l}_{00\leftarrow 10}$ does the
opposite.

We obtain the steady-state value of $\brho$, denoted $\brhobar$, by solving
$d\brhobar/dt=\mathcal{M}\brhobar=0$. Using techniques described in detail in
Refs.~\cite{hershfield93,eto97,kiesslich03}, we define current matrices $J^{tr}$
and $J^{br}$ for the top-right and bottom-right leads and apply them to $\brhobar$
to calculate the average currents $\langle I_{t(b)}\rangle$ and the correlator
$\langle \It(\tau)\Ib(0)\rangle$~\cite{matrices}. The cross-spectral density in
the low-frequency limit is then given by
\begin{equation}
\six= 2\int_{-\infty}^{\infty} \left[\langle\It(\tau)\Ib(0)\rangle - \langle \It
\rangle \langle \Ib \rangle \right] d\tau.
\end{equation}

Simulation results for noise cross-correlation $\six$ as a function of plunger
gate voltages are shown in Fig.~2(b), with all parameters of the model extracted
from experiment: $U=60~\mueV$, $\te=280~\mk$, $\Gamma^{tl}= \Gamma^{tr} =
1.5\times 10^{10}~\mathrm{s}^{-1}$, and $\Gamma^{bl}=\Gamma^{br}=
7.2\times10^{9}~\mathrm{s}^{-1}$. The $\Gamma^i$ were estimated from the zero-bias
conductance peak height using Eq.~6.3 from Ref.~\cite{beenakker91} and assuming
symmetric left and right barriers. The simulation shows the characteristic
quadrupole pattern of positive and negative cross-correlation observed
experimentally. We note that the model underestimates $\six$ by roughly a factor
of two. This may be due to transport through processes not accounted for in the
model. Elastic cotunneling may be present since the $\Gamma^i$ are comparable to
$\kb\te/\hbar$. Furthermore, since the voltage-bias energy $|eV_{t(b)}|$ is
greater than the level spacing $\Delta_{t(b)}$, transport may occur via multiple
levels~\cite{belzig05,cottet04} and via inelastic
cotunneling~\cite{sukhorukov01,onac06}.

Some intuition for how Coulomb interactions affect noise cross-correlation in this
device geometry can be gained from simple considerations of energy level positions
in the space of plunger gate voltages, illustrated in Fig.~3. With both dots
positioned near Coulomb blockade peaks, the fluctuations by one in the electron
number of each dot, caused by the sequential tunneling of electrons through that
dot, cause the energy level of the other dot to fluctuate between two values
separated by $U$. These fluctuations can raise and lower the level across the
chemical potential in one of the dot's leads, strongly affecting either the
tunnel-in rate (from the left, for the case illustrated in Fig.~3) or the
tunnel-out rate (to the right) of that dot. Specifically, the rate of the
``$U$-sensitive'' process in each dot fluctuates between a slow value (red arrow),
suppressed significantly from $\Gamma^i$, and a fast value (green arrow), close to
$\Gamma^i$. Since the $\Gamma^i$ for the left and right barriers of each dot are
roughly equal, the $U$-sensitive process becomes the rate-limiting process for
transport through the dot when its rate is suppressed.

The $U$-sensitive processes correlate transport through the two dots. In region
(b) of Fig.~3, for instance, where $\six$ is negative, the $U$-sensitive process
in each dot is tunneling-out. Here, as well as in (c), where the $U$-sensitive
process in each dot is tunneling-in, the two $U$-sensitive processes compete: the
occurrence of one suppresses the rate of the other, leading to negative $\six$.
Conversely, in region (a) [(d)], where $\six$ is positive, the $U$-sensitive
process in the top [bottom] dot is tunneling-out, and in the bottom [top] dot it
is tunneling-in. Here, the $U$-sensitive processes cooperate: the occurrence of
one lifts the suppression of the other, leading to positive $\six$.

\begin{figure}[t]
\center \label{figure4}
\includegraphics[width=3.25in]{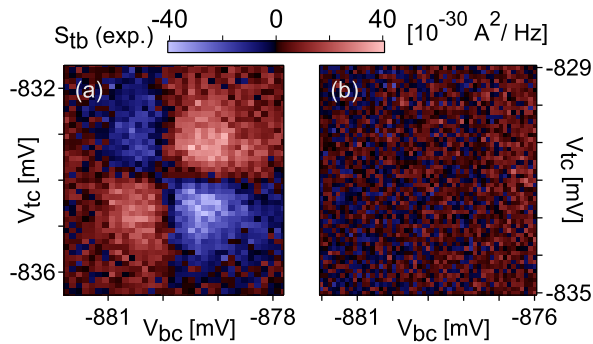}
\caption{\footnotesize{(a) Measured $\six$ near a honeycomb vertex, with opposite
biases $\vt=-\vb=-100~\muV$. Note that the pattern is reversed from Fig.~2(a):
negative cross-correlation (blue) is now found in the upper-left and lower-right
regions, while positive cross-correlation (red) is now found in the lower-left and
upper-right. (b) Measured $\six$ near a honeycomb vertex, with $\vt=\vb=0$.
Cross-correlation of noise vanishes at zero bias, though the noise in each dot is
finite.}}
\end{figure}

The arguments above also apply when one or both biases are reversed. When both are
reversed, we find both experimentally and in the model that the same
cross-correlation pattern as in Fig.~2 appears (not shown). When only one of the
biases is reversed, we find both experimentally [as shown in Fig.~4(a)] and in the
model that the pattern reverses sign. In the absence of any bias,
cross-correlation vanishes both experimentally [as shown in Fig.~4(b)] and in the
model, despite the fact that noise in the individual dots remains finite (as
seen in Fig.~1(c))

In conclusion, we have observed gate-controlled sign reversal of current noise
cross-correlation in two capacitively coupled quantum dots. Our ability to tune
the parameters governing transport through the localized states allows us to
observe a distinctive noise signature of Coulomb interaction between these states
that agrees with the predictions of a simple model.

We thank N.~J.~Craig for device fabrication and M.~Eto, W.~Belzig, C.~Bruder,
E.~Sukhorukov, and L.~Levitov for valuable discussions. We acknowledge support
from the NSF through the Harvard NSEC, PHYS 01-17795, DMR-05-41988, DMR-0501796,
as well as support from NSA/DTO and Harvard University.


\begin{thebibliography}{31}
\bibitem[{\citenamefont{MATRICESGOHERE}()}]{matrices}
Following the approach of Refs.~\cite{hershfield93,eto97,kiesslich03}, we define
the current matrices as follows: $J^{tr}_{mn,m^\prime
n^\prime}=|e|\delta_{nn^\prime}(m-m^\prime)\rate^r_{mn\leftarrow m^\prime
n^\prime}$ and $J^{br}_{mn,m^\prime
n^\prime}=|e|\delta_{mm^\prime}(n-n^\prime)\rate^r_{mn\leftarrow m^\prime
n^\prime}$.  The average currents are given by $\langle
I_{t(b)}\rangle=\sum_{i}\left(J^{t(b)r}\brhobar\right)_{i}$, and the low-frequency
correlator by $\langle \It(\tau)\Ib(0) \rangle=\sum_{i}\left(\theta(\tau)\Jtr
e^{\Mmat \tau}\Jbr \brhobar+\theta(-\tau)\Jbr e^{\Mmat \tau}\Jtr
\brhobar\right)_{i}$, where $\theta$ is the Heaviside step function.
\end{thebibliography}
\end{document}